\documentclass[prx,twocolumn,superscriptaddress,floatfix,nopacs,longbibliography]{revtex4-1}
\usepackage{graphicx,amsfonts,amssymb,amsmath,hyperref,enumerate}

\newif\ifhyper
\hypertrue
\ifhyper
\hypersetup{
   citecolor = {green},
   colorlinks = {true}, 
   urlcolor = {blue} 
}

\newcommand{\w}{{\omega}}

\begin{document} 

\title{Hybrid Quantum Investment Optimization with Minimal Holding Period}

\author{Samuel Mugel}
\affiliation{Multiverse Computing, Centre for Social Innovation, 192 Spadina Ave, Suite 412, Toronto M5T 2C2, Canada}

\author{Mario Abad}
\affiliation{Bankia Asset Management, Paseo de la Castellana, 189 28046 Madrid, Spain}

\author{Miguel Bermejo}
\affiliation{Bankia Innovation \& Cybersecurity, Paseo de la Castellana, 189 28046 Madrid, Spain}

\author{Javier S\'anchez}
\affiliation{Bankia Innovation \& Cybersecurity, Paseo de la Castellana, 189 28046 Madrid, Spain}

\author{Enrique Lizaso} 
\affiliation{Multiverse Computing, Paseo de Miram\'on 170, E-20014 San Sebasti\'an, Spain}

\author{Rom\'an Or\'us}
\affiliation{Multiverse Computing, Paseo de Miram\'on 170, E-20014 San Sebasti\'an, Spain}
\affiliation{Donostia International Physics Center, Paseo Manuel de Lardizabal 4, E-20018 San Sebasti\'an, Spain}
\affiliation{Ikerbasque Foundation for Science, Maria Diaz de Haro 3, E-48013 Bilbao, Spain}

\affiliation{\textit {Corresponding author: \href{mailto:roman.orus@dipc.org}{roman.orus@dipc.org}}}

\begin{abstract}
\begin{center}
\textbf{\abstractname}
\end{center}
In this paper we propose a hybrid quantum-classical algorithm for dynamic portfolio optimization with minimal holding period. Our algorithm is based on sampling the near-optimal portfolios at each trading step using a quantum processor, and efficiently post-selecting to meet the minimal holding constraint. We found the optimal investment trajectory in a dataset of 50 assets spanning a one year trading period using the D-Wave 2000Q processor. Our method is remarkably efficient, and produces results much closer to the efficient frontier than typical portfolios. Moreover, we also show how our approach can easily produce trajectories adapted to different risk profiles, as typically offered in financial products. Our results are a clear example of how the combination of quantum and classical techniques can offer novel  valuable tools to deal with real-life problems, beyond simple toy models, in current NISQ quantum processors.  
\end{abstract}

\maketitle

\emph{Introduction.---} The field of quantum computing is living an unprecedented expansion due to recent experimental advances. {Quantum technologies as a whole \cite{Q1, Q2, Q3, Q4, Q5, Q6, Q7, Q8} are being benefited from recent results both on quantum computer implementations \cite{Q21, Q22, Q23, Q24, Q25, Q26, Q27} as well as on quantum communications \cite{Q31, Q32, Q33, Q34, Q35, Q36, Q37}.} Due to this, people have started to think seriously about industrial applications of quantum computers \cite{MacQuarrie2020}. Among the different verticals, finance is among the most promising ones, given the ubiquitousness of intractable mathematical problems. For a detailed description of applications of quantum computing in finance, see Ref. \cite{Orus2018}. Among these applications, one of the most prominent is quantum optimization. There are many important optimization problems in finance which can be solved more efficiently using quantum computing. See Refs. \cite{Mugel2020a, Rosenberg2016,Elsokkary2017a,Grant2020,Cohen2020} for some examples. In this setting, the most paradigmatic optimization problem in finance is that of portfolio optimization, both in its static and dynamic versions. 

Our aim in this work is to solve the dynamic portfolio optimization problem. An issue that investors often face is that short term investments tend to be taxed much higher than long term investments. It is common for investors to impose a minimal holding period, preventing any purchased asset from being sold before a predetermined period of time. We build upon the work from Ref. \cite{Mugel2020}, and demonstrate an efficient post-selection protocol to impose the minimal holding constraint. Similarly to Ref. \cite{Mugel2020}, we further impose that investors must invest in integer bundles, as is typically the case for exchange-traded funds (ETF) shares. 

\emph{Financial model.---} Our aim is to find the best investment trajectory for an investor, given a level of risk that they want to take. According to Modern Portfolio Theory, the optimal investment at a defined level of risk is the one which maximizes profit \cite{Singleton2018}. The portfolio's risk -- or \emph{volatility} -- is computed from the assets' covariance matrix. The ratio of returns to risk is the \emph{Sharpe ratio}, which is our metric for comparing investments. {These are defined mathematically in what follows.}

Let us define $\w_{tn}$, the fraction of the total budget invested in asset $n$ at time $t$. The optimal holding trajectory $\w_{tn}$ minimizes the Modern Portfolio Theory cost function:
\begin{equation}
\label{eq:bare_cost_function}
H_0 =  \sum_{t} 
 -\mu_t^T \w_t
 + \frac{\gamma}{2} \w_t^T \Sigma_t \w_t.
\end{equation}
Here, $\mu_t$ is the vector of the logarithmic returns at time $t$ and $\Sigma_t$ is the covariance matrix. The \emph{risk aversion} $\gamma$ controls the portfolios penalty for risk. This determines the amount of risk an investor is willing to take. The forecasted returns and covariance matrices can be deduced from the stocks' {prices}. This is detailed, for instance, in Ref. \cite{Mugel2020}. {Following this notation, the Sharpe ratio is given by 
\begin{equation}
{\rm Sharpe} \equiv \frac{\sum_{t} \mu_t^T \w_t}{\sqrt{ \sum_t \w_t^T \Sigma_t \w_t}}, 
\end{equation}
which, as we said, is nothing but the ration between {the total return} (numerator) and volatility (denominator).}

At any point in time, we would like the entire available budget to be invested. This constrains the holdings $\w_t$ to be normalized at any time $t$. We enforced this by penalizing portfolios which did not respect this constraint. The cost function we optimized is therefore
\begin{equation}
\label{eq:cost_function}
H = H_0 + \sum_t \rho \left(\sum_n \w_{tn} - 1 \right)^2,
\end{equation}
where the Lagrange multiplier $\rho$ is an hyperparameter of the model. Note also that Eq. \eqref{eq:cost_function} can be written as:
\begin{equation}
\label{eq:diagonal_hamiltonian}
H = \sum_t h_t,
\end{equation}
for some $h_t, t \in \{0, N_t\}$. This means that the optimal investment at time $t$ is independent of our investment history, as long as no more constraints are included that correlate trading times among each other. In such a situation it is therefore sufficient to minimize $h_t$ at every time $t$ to compute the optimal holdings $\w_t$. This is a good approximation, for instance, for the trading investment funds of an everyday investor, for which transaction costs and market impact may be negligible. 

We assume shares can only be sold in large bundles. These constraints imply that our objective variables -- the instantaneous asset holdings  $\w_{tn}$ -- are integer variables. At any trading time, we can encode the qubit values to our qubits using a binary encoding:
\begin{equation}
\label{eq:encoding}
\w_{n} = \frac{1}{K} \sum_{q=0}^{N_q-1}  2^q x_{n, q},
\end{equation}
where $x_{n, q} \in \{0,1\}$ is the readout value of the $q^\text{th}$ qubit assigned to institution $n$, {and $K$ is the total investment.}

This encoding has several consequences. First, the holdings are always bound by: $\w_{tn} \in [0, 1]$. Since $\w_{tn} \ge 0$, investors are not given the option to sell short securities. Allowing short-selling could be an interesting extension for this problem. Second, this encoding allows investors to split their total investment in a maximum of $K$ bundles. This discretization is part of the reason why this problem is so hard to solve on classical computers \cite{Mugel2020}. {From a practical perspective, using the bit variables the cost function can then be written as 
\begin{equation}
H = x^T Q x, 
\end{equation}
with $x$ a vector of bit variables, and $Q$ a matrix of real numbers. In the language of combinatorial optimization, this is a quadratic unconstrained binary optimization (QUBO) problem, and is the natural input for the D-Wave quantum annealer.} 

In Eq. \eqref{eq:encoding}, we have introduced the \emph{bit depth} $N_q$, the number of qubits which encode the $n^\text{th}$ asset holdings. Typically, we would choose $N_q$ such that $2^{N_q} -1 \geq K$. This choice allows the investor to invest their entire budget into a single asset. We may, however, want to impose a diversification constraint. Typically, large financial institutions are not allowed to invest more than $5\%$ of their total budget in any single asset. Choosing $N_q$ such that $2^{N_q} -1 = 0.05 K$ naturally implements this diversification constraint.

\begin{figure}[t]
  \centering
      \includegraphics[width=0.5\textwidth]{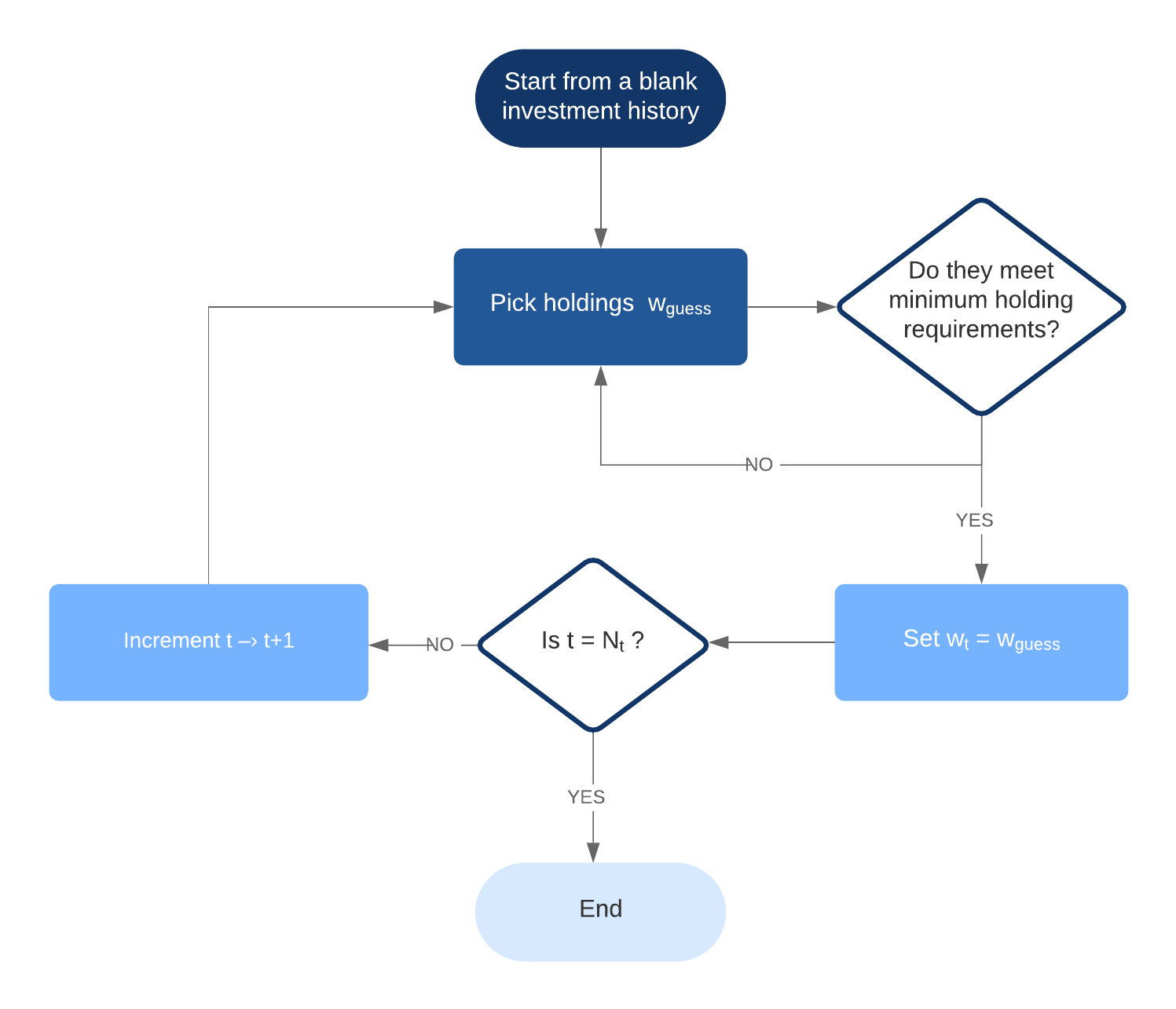}
  \caption{(Colour online) Flow chart detailing the post-selection algorithm used to efficiently eliminate trajectories which do not meet the minimum 7 day holding period.}
  \label{fig:flow_chart}
\end{figure}

\begin{figure}[t]
  \centering
      \includegraphics[width=0.5\textwidth]{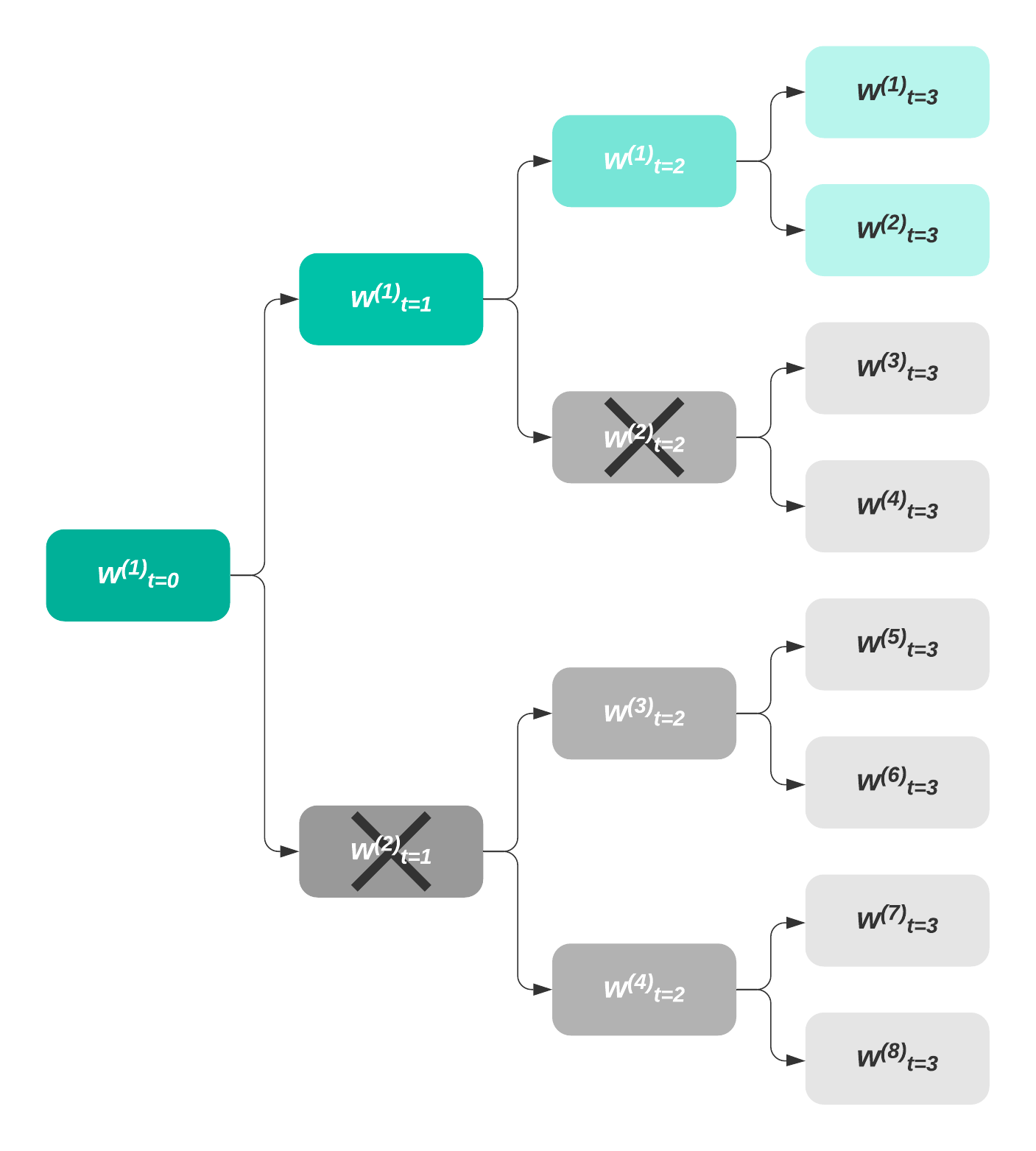}
  \caption{(Colour online) Candidate investment trajectories are efficiently ruled out by the post-selection algorithm. Node $\w^{(i)}_t$ represents the $i^\text{th}$ candidate holdings at time $t$. Green nodes meet the minimum holding period, while grey nodes do not. When the constraint is not met at time $t$, the node is crossed out and all resulting investment trajectories are eliminated.}
  \label{fig:minimum_holding_constraint}
\end{figure}

\emph{Minimum holding period.---} The \emph{holding period} is the amount of time which elapses between an investment's purchase and its sale (or sale of a security). Because long-term gains are taxed more favourably than short-term gains, it is common to demand that investments meet a minimum holding period. In this study, we imposed a minimal holding period of seven days, which is a natural timeframe for some realistic banking products.

Here we imposed this constraint by post-selecting investment trajectories which respected this condition. The number of possible trajectories grows exponentially $O\left(N_a^{K \times N_t}\right)$. It is therefore computationally prohibitive to successively consider each trajectory and verify if they meet the minimal 7 day holding period.

Instead, we build the investment trajectory iteratively, following the flow chart presented Fig. \ref{fig:flow_chart}. At every time $t$, we choose holdings $\w_\text{guess}$ for that instant. {We compute them by sampling at each time $t$ the low-energy subspace of  $h_t$ in Eq. \eqref{eq:diagonal_hamiltonian} using, in our case, the D-Wave machine. the fact that the problem can be decoupled into separate times $t$ allows us to tackle more assets at every given time with the quantum annealer. The sampled instantaneous portfolios are considered in order of decreasing Sharpe ratio. As customary in finance, a large Sharpe ratio is an indicator of a good-quality portfolio, since it implies that the return is large for the risk being assumed. We note, however, that other figures of merit could be used, such as the returns (the larger the better), or the volatility (the lower the better). In our case, though we choose to work with the Sharpe ratio because of the afore-mentioned reasons.} We only retain candidates which fulfil the minimum holding constraint.

{By following iteratively, the algorithm implements a tree-like selection and efficiently rules out trajectories which do not meet the minimal 7-day holding period, as is illustrated in Fig. \ref{fig:minimum_holding_constraint}.} Note that at any time $t$, there always exists at least one solution which meets the minimal 7-day holding period: the solution $\w_t = \w_{t-1}$.

Alternatively, we could include this constraint by penalizing in Eq. \eqref{eq:cost_function} trajectories which do not meet the minimum holding period. The corresponding penalty term would then correlate different time steps between each other, such that the cost function would not have the separable form of Eq. (\ref{eq:diagonal_hamiltonian}). In such a case, the problem would become even more computationally intractable, and a global optimization strategy over all possible trading times would be necessary from the very beginning. It is therefore remarkable that our simple approach, based on quantum sampling and post-selection, produces such high-quality portfolios in an extremely efficient manner. 

\emph{Dimensional reduction.---} The number of objective variables in this problem is proportional to the number of assets, multiplied by the bit depth. For problems of commercial value, this can be very large. Because current quantum resources in Noisy Intermediate Scale Quantum (NISQ) devices are limited, a good option is to apply dimensional reduction techniques. Reducing the space of solutions searched also makes it easier for the optimization routine to converge to the global minimum.

In Ref. \cite{Mugel2020}, authors described a method to cluster assets based on their time series' similarity. In the following, we apply this method, and discard all except for the best asset in each cluster (based on its historical Sharp ratio). {The dimensional reduction strategy follows from applying first a Hodrick-Prescott smoothing, to extract data trends, and then computing the euclidean distance between different pairs of trends of assets. Thanks to this, we can perform a clustering of the assets into clusters of assets of similar behavior.} This dimensional reduction strategy also allows us to significantly lower the portfolio's risk, by diversifying our investment among maximally uncorrelated assets. Moreover, our target investors typically select their investment risk category. To meet this risk requirement, we performed a pre-selection among available assets. We computed each assets' historical volatility, and discarded any asset which significantly exceeded the agreed risk level. This allows us to construct portfolios for investors with different risk profiles: high-risk, medium-risk, low-risk, and so forth. 

In practice, we determined the optimal number of clusters $N_c$ by studying each clusters' variance. In the studied data, we found that the mean clusters' variance decreased exponentially with the number of clusters. For $N_c>7$, the clusters' variance did not significantly decrease, indicating that 7 clusters capture almost all of the system's variations. We therefore set $N_c=7$ in our calculations. 

\begin{figure}[t]
  \centering
    \includegraphics[width=0.5\textwidth]{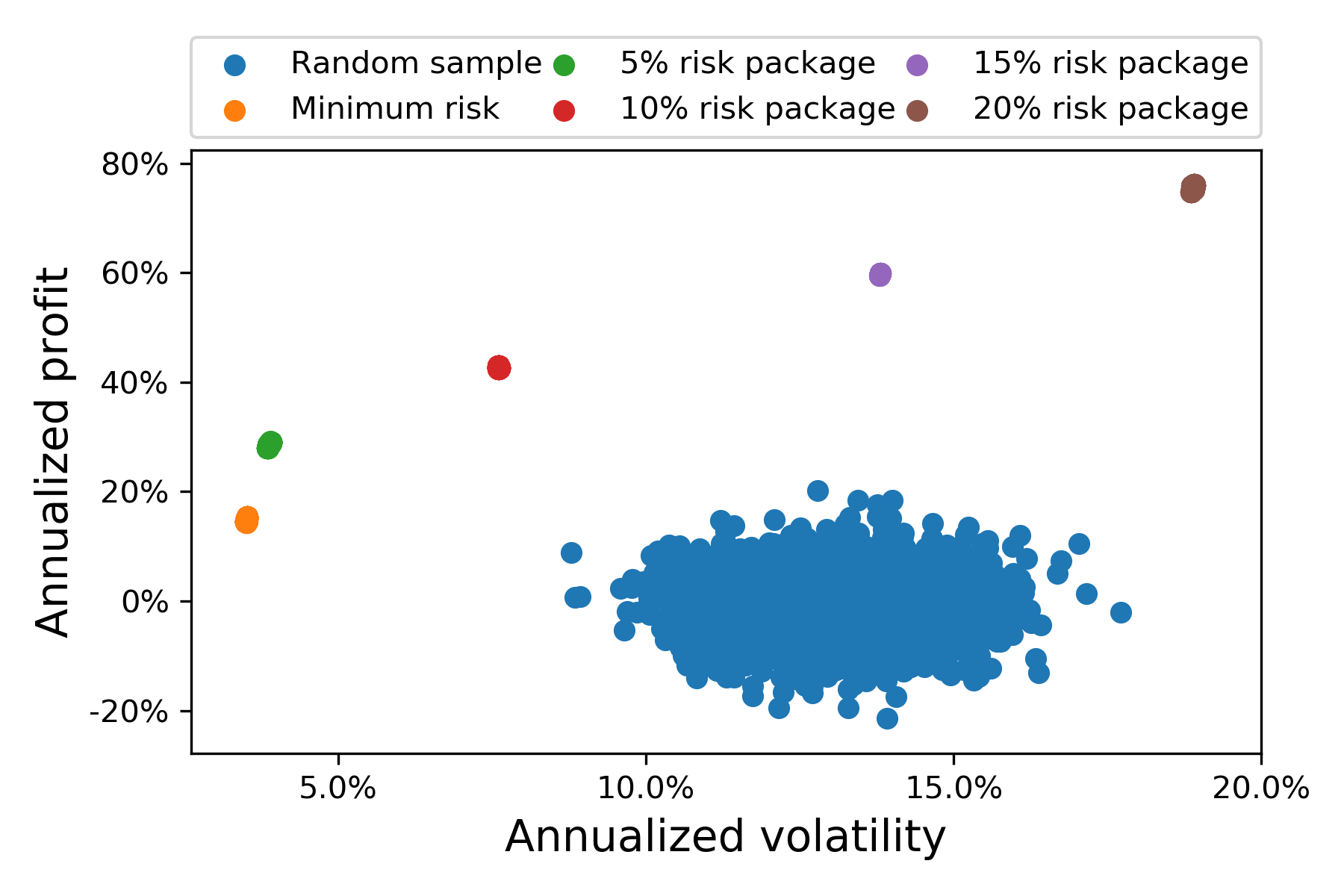}
    \caption{(Colour online) Investment trajectories, chosen among 50 assets spanning May the 31$^\text{st}$ 2019 to May the 31$^\text{st}$ 2020. The blue dots represent randomly selected trajectories. The coloured dots are investment trajectories with different levels of risk obtained using our quantum optimization toolbox.}
    \label{fig:portfolio_trajectories}
\end{figure}

\emph{Results.---} Following the protocol detailed in previous sections, {based on looking at optimal Sharpe ratios of portfolios computed with the D-Wave quantum annealer,} we found the optimal investment trajectory between May the 31$^\text{st}$ 2019 and May the 31$^\text{st}$ 2020 among a total of 50 international assets and indices. Daily transactions were allowed, but investments had to meet a minimum seven day holding period. We allowed investments to be split into a maximum of $K=5$ bundles, and set the bit depth to $N_q=2$.

\begin{figure*}
  \centering
      \includegraphics[width=\textwidth]{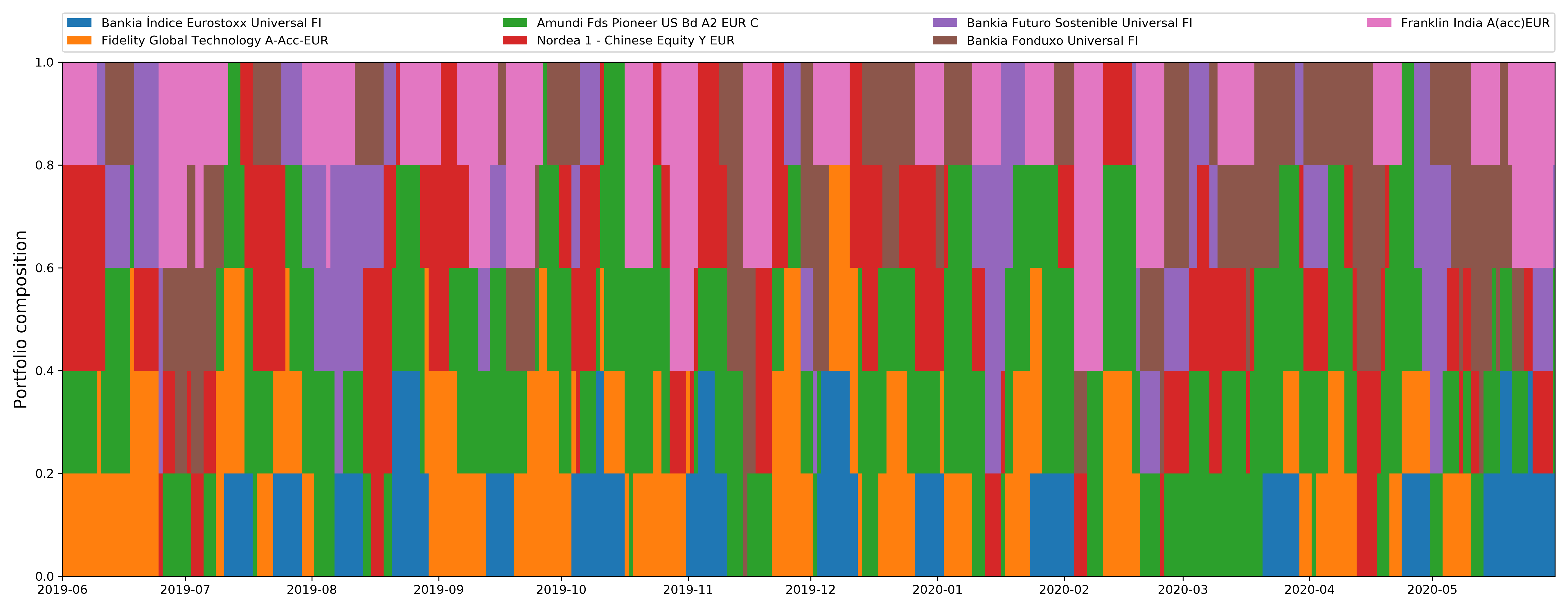}
  \caption{(Colour online) The optimal investment trajectory between May the 31$^\text{st}$ 2019 and May the 31$^\text{st}$ 2020 among considered assets for an investor wanting to take $\leq15 \%$ risk. As can be seen from Fig.\ \ref{fig:portfolio_trajectories}, this portfolio provided $\approx 60\%$ annualized return on investment (purple dot in the Fig. \ref{fig:portfolio_trajectories}). {Data for this calculation was obtained from daily prices of international assets and indices as explained in the text, which are public and can be obtained from e.g. Bloomberg, Yahoo Finance, and/or Morningstar databases.}}
  \label{fig:portfolio_holdings}
\end{figure*}

We found the optimal portfolios by minimizing Eq. \eqref{eq:cost_function} on the D-Wave 2000Q processor. Computed investment trajectories for five risk packages (minimum risk, $5\%$, $10\%$, $15\%$, and $20\%$ annual  risk) are plotted in Fig. \ref{fig:portfolio_trajectories}. Also plotted are 1000 random investment trajectories which meet the minimum seven day holding period. These give us some intuition about the space of allowed portfolios. We see that allowed portfolios tend to exist between $-20\%$ and $20\%$ annualized profit, and between $10\%$ and $17\%$ risk (as measured by the annualized volatility). It is clearly visible that the investment trajectories obtained using our method based on quantum optimization, dimensional reduction and post-selection, tend to gather much closer to the efficient frontier in portfolio space than those obtained by investing at random.

In Fig. \ref{fig:portfolio_holdings}, we show an example investment trajectory. This corresponds to the $15 \%$ risk package (purple dot in Fig. \ref{fig:portfolio_trajectories}), which yields $\approx 60\%$ return on investment. We can see from this figure that the $N_c=7$ most profitable assets which meet the $15 \%$ risk limit are Bankia Fonduxo Universal FI, Bankia Futuro Sostenible Universal FI, Bankia Indice Eurostoxx Universal FI, Amundi Fds Pioneer US Bd A2 EUR C, Fidelity Global Technology A-Acc-EUR, Nordea 1 - Chinese Equity Y EUR, and Franklin India A(acc)EUR. As can be read off from Fig. \ref{fig:portfolio_holdings}, the optimal investment on May the 31$^\text{st}$ 2019 meeting these constraints is: $1/5^\text{th}$ in Fidelity Global Technology A-Acc-EUR, $1/5^\text{th}$ in Amundi Fds Pioneer US Bd A2 EUR C, $1/5^\text{th}$ in Franklin India A(acc)EUR, and $2/5^\text{th}$ in Nordea 1 - Chinese Equity Y EUR. The everyday composition of the trading trajectory can be read form the chart, and it always fulfils by construction the minimum holding period. 

{Finally, we would like to discuss about the computational running time and quality of the computed portfolios, in comparison with other classical and quantum optimization strategies. The running time to compute the optimal portfolios in Fig.\ref{fig:portfolio_trajectories} using our strategy of combining D-Wave and post-selection was of \emph{just a few minutes}, for daily portfolios of 50 assets. For comparison, other quantum optimization strategies, such as Variational Quantum Eigensolvers (VQE) on universal quantum processors are highly limited at the moment and can only handle very small portfolios with typically 2-3 assets at most \cite{Mugel2020}. Compared to classical solvers, our method is way faster than brute-force classical search, which scales exponentially fast in the number of qubits and quickly hits astronomical numbers (the problem is clearly intractable in this sense for 50 assets). Compared to other solvers, D-Wave quantum annealing is known to be faster and allow larger portfolios than solvers such as Gekko, see Table IV in Ref.\cite{Mugel2020}, which equally applies in our case.}

\emph{Conclusion.---} Here we have proposed a hybrid quantum-classical algorithm for dynamic portfolio optimization with minimal holding period. Our algorithm is based on quantum sampling of near-optimal portfolios at each trading step, combined with dimensional reduction and post-selection. We tested our idea using quantum optimization with the D-Wave 2000Q processor, and real data from 50 assets during one year. Our study shows that the method is remarkably efficient and produces {in few minutes} results close to the optimal efficient frontier in portfolio space, much better than typical random portfolios. We also showed how our approach can be used to produce trajectories corresponding to different risk profiles, in accordance with typical products offered by financial institutions to investors. Our results are a clear example of how the combination of quantum and classical techniques can provide novel and valuable methods to deal with real-life financial data in today's NISQ quantum processors. {Moreover, we believe that the use of distributed quantum computer architectures, such as those discussed in Ref.\cite{Q3}, could accelerate even more the implementation of this type of quantum computing algorithms to solve real-life financial mathematical problems.} 

\bigskip 

{\bf Acknowledgments:} We wish to thank Ignacio Cea from Bankia, as well as Carlos Ceruelo and \'Angel S\'anchez from Everis, for enabling this fascinating project. We thank Christophe Jurczak, Pedro Luis Uriarte, Pedro Mu{\~n}oz-Baroja, Joseba Sagastigordia, Creative Destruction Lab, BIC-Gipuzkoa, DIPC, Ikerbasque, and Basque Government for constant support. We also acknowledge the hard work and constant feedback from Multiverse's fantastic technical team, Francesco Benfenati, Gianni Del Bimbo, Rodrigo Hernandez, Beinat Mencia, Samuel Palmer, and Serkan Sahin.

\bibliography{bibliography}

\begin{thebibliography}{31}%
\makeatletter
\providecommand \@ifxundefined [1]{%
 \@ifx{#1\undefined}
}%
\providecommand \@ifnum [1]{%
 \ifnum #1\expandafter \@firstoftwo
 \else \expandafter \@secondoftwo
 \fi
}%
\providecommand \@ifx [1]{%
 \ifx #1\expandafter \@firstoftwo
 \else \expandafter \@secondoftwo
 \fi
}%
\providecommand \natexlab [1]{#1}%
\providecommand \enquote  [1]{``#1''}%
\providecommand \bibnamefont  [1]{#1}%
\providecommand \bibfnamefont [1]{#1}%
\providecommand \citenamefont [1]{#1}%
\providecommand \href@noop [0]{\@secondoftwo}%
\providecommand \href [0]{\begingroup \@sanitize@url \@href}%
\providecommand \@href[1]{\@@startlink{#1}\@@href}%
\providecommand \@@href[1]{\endgroup#1\@@endlink}%
\providecommand \@sanitize@url [0]{\catcode `\\12\catcode `\$12\catcode
  `\&12\catcode `\#12\catcode `\^12\catcode `\_12\catcode `\%12\relax}%
\providecommand \@@startlink[1]{}%
\providecommand \@@endlink[0]{}%
\providecommand \url  [0]{\begingroup\@sanitize@url \@url }%
\providecommand \@url [1]{\endgroup\@href {#1}{\urlprefix }}%
\providecommand \urlprefix  [0]{URL }%
\providecommand \Eprint [0]{\href }%
\providecommand \doibase [0]{http://dx.doi.org/}%
\providecommand \selectlanguage [0]{\@gobble}%
\providecommand \bibinfo  [0]{\@secondoftwo}%
\providecommand \bibfield  [0]{\@secondoftwo}%
\providecommand \translation [1]{[#1]}%
\providecommand \BibitemOpen [0]{}%
\providecommand \bibitemStop [0]{}%
\providecommand \bibitemNoStop [0]{.\EOS\space}%
\providecommand \EOS [0]{\spacefactor3000\relax}%
\providecommand \BibitemShut  [1]{\csname bibitem#1\endcsname}%
\let\auto@bib@innerbib\@empty
\bibitem [{\citenamefont {Gyongyosi}\ and\ \citenamefont
  {Imre}(2019{\natexlab{a}})}]{Q1}%
  \BibitemOpen
  \bibfield  {author} {\bibinfo {author} {\bibfnamefont {Laszlo}\ \bibnamefont
  {Gyongyosi}}\ and\ \bibinfo {author} {\bibfnamefont {Sandor}\ \bibnamefont
  {Imre}},\ }\bibfield  {title} {\enquote {\bibinfo {title} {A survey on
  quantum computing technology},}\ }\href {\doibase
  https://doi.org/10.1016/j.cosrev.2018.11.002} {\bibfield  {journal} {\bibinfo
   {journal} {Computer Science Review}\ }\textbf {\bibinfo {volume} {31}},\
  \bibinfo {pages} {51--71} (\bibinfo {year} {2019}{\natexlab{a}})}\BibitemShut
  {NoStop}%
\bibitem [{\citenamefont {Gyongyosi}\ \emph {et~al.}(2018)\citenamefont
  {Gyongyosi}, \citenamefont {Imre},\ and\ \citenamefont {Nguyen}}]{Q2}%
  \BibitemOpen
  \bibfield  {author} {\bibinfo {author} {\bibfnamefont {Laszlo}\ \bibnamefont
  {Gyongyosi}}, \bibinfo {author} {\bibfnamefont {Sandor}\ \bibnamefont
  {Imre}}, \ and\ \bibinfo {author} {\bibfnamefont {Hung~Viet}\ \bibnamefont
  {Nguyen}},\ }\bibfield  {title} {\enquote {\bibinfo {title} {A survey on
  quantum channel capacities},}\ }\href {\doibase 10.1109/COMST.2017.2786748}
  {\bibfield  {journal} {\bibinfo  {journal} {IEEE Communications Surveys
  Tutorials}\ }\textbf {\bibinfo {volume} {20}},\ \bibinfo {pages} {1149--1205}
  (\bibinfo {year} {2018})}\BibitemShut {NoStop}%
\bibitem [{\citenamefont {Gyongyosi}\ and\ \citenamefont {Imre}(2021)}]{Q3}%
  \BibitemOpen
  \bibfield  {author} {\bibinfo {author} {\bibfnamefont {Laszlo}\ \bibnamefont
  {Gyongyosi}}\ and\ \bibinfo {author} {\bibfnamefont {Sandor}\ \bibnamefont
  {Imre}},\ }\bibfield  {title} {\enquote {\bibinfo {title} {Scalable
  distributed gate-model quantum computers},}\ }\href {\doibase
  10.1038/s41598-020-76728-5} {\bibfield  {journal} {\bibinfo  {journal}
  {Scientific Reports}\ }\textbf {\bibinfo {volume} {11}},\ \bibinfo {pages}
  {5172} (\bibinfo {year} {2021})}\BibitemShut {NoStop}%
\bibitem [{\citenamefont {MacQuarrie}\ \emph
  {et~al.}(2020{\natexlab{a}})\citenamefont {MacQuarrie}, \citenamefont
  {Simon}, \citenamefont {Simmons},\ and\ \citenamefont {Maine}}]{Q4}%
  \BibitemOpen
  \bibfield  {author} {\bibinfo {author} {\bibfnamefont {Evan~R.}\ \bibnamefont
  {MacQuarrie}}, \bibinfo {author} {\bibfnamefont {Christoph}\ \bibnamefont
  {Simon}}, \bibinfo {author} {\bibfnamefont {Stephanie}\ \bibnamefont
  {Simmons}}, \ and\ \bibinfo {author} {\bibfnamefont {Elicia}\ \bibnamefont
  {Maine}},\ }\bibfield  {title} {\enquote {\bibinfo {title} {The emerging
  commercial landscape of quantum computing},}\ }\href {\doibase
  10.1038/s42254-020-00247-5} {\bibfield  {journal} {\bibinfo  {journal}
  {Nature Reviews Physics}\ }\textbf {\bibinfo {volume} {2}},\ \bibinfo {pages}
  {596--598} (\bibinfo {year} {2020}{\natexlab{a}})}\BibitemShut {NoStop}%
\bibitem [{\citenamefont {Gyongyosi}\ and\ \citenamefont
  {Imre}(2020{\natexlab{a}})}]{Q5}%
  \BibitemOpen
  \bibfield  {author} {\bibinfo {author} {\bibfnamefont {Laszlo}\ \bibnamefont
  {Gyongyosi}}\ and\ \bibinfo {author} {\bibfnamefont {Sandor}\ \bibnamefont
  {Imre}},\ }\bibfield  {title} {\enquote {\bibinfo {title} {Circuit depth
  reduction for gate-model quantum computers},}\ }\href {\doibase
  10.1038/s41598-020-67014-5} {\bibfield  {journal} {\bibinfo  {journal}
  {Scientific Reports}\ }\textbf {\bibinfo {volume} {10}},\ \bibinfo {pages}
  {11229} (\bibinfo {year} {2020}{\natexlab{a}})}\BibitemShut {NoStop}%
\bibitem [{\citenamefont {Gyongyosi}\ and\ \citenamefont
  {Imre}(2020{\natexlab{b}})}]{Q6}%
  \BibitemOpen
  \bibfield  {author} {\bibinfo {author} {\bibfnamefont {Laszlo}\ \bibnamefont
  {Gyongyosi}}\ and\ \bibinfo {author} {\bibfnamefont {Sandor}\ \bibnamefont
  {Imre}},\ }\bibfield  {title} {\enquote {\bibinfo {title} {Optimizing
  high-efficiency quantum memory with quantum machine learning for near-term
  quantum devices},}\ }\href {\doibase 10.1038/s41598-019-56689-0} {\bibfield
  {journal} {\bibinfo  {journal} {Scientific Reports}\ }\textbf {\bibinfo
  {volume} {10}},\ \bibinfo {pages} {135} (\bibinfo {year}
  {2020}{\natexlab{b}})}\BibitemShut {NoStop}%
\bibitem [{\citenamefont {Gyongyosi}(2020)}]{Q7}%
  \BibitemOpen
  \bibfield  {author} {\bibinfo {author} {\bibfnamefont {Laszlo}\ \bibnamefont
  {Gyongyosi}},\ }\bibfield  {title} {\enquote {\bibinfo {title} {Quantum state
  optimization and computational pathway evaluation for gate-model quantum
  computers},}\ }\href {\doibase 10.1038/s41598-020-61316-4} {\bibfield
  {journal} {\bibinfo  {journal} {Scientific Reports}\ }\textbf {\bibinfo
  {volume} {10}},\ \bibinfo {pages} {4543} (\bibinfo {year}
  {2020})}\BibitemShut {NoStop}%
\bibitem [{\citenamefont {Gyongyosi}\ and\ \citenamefont
  {Imre}(2019{\natexlab{b}})}]{Q8}%
  \BibitemOpen
  \bibfield  {author} {\bibinfo {author} {\bibfnamefont {Laszlo}\ \bibnamefont
  {Gyongyosi}}\ and\ \bibinfo {author} {\bibfnamefont {Sandor}\ \bibnamefont
  {Imre}},\ }\bibfield  {title} {\enquote {\bibinfo {title} {Dense quantum
  measurement theory},}\ }\href {\doibase 10.1038/s41598-019-43250-2}
  {\bibfield  {journal} {\bibinfo  {journal} {Scientific Reports}\ }\textbf
  {\bibinfo {volume} {9}},\ \bibinfo {pages} {6755} (\bibinfo {year}
  {2019}{\natexlab{b}})}\BibitemShut {NoStop}%
\bibitem [{\citenamefont {Arute}\ \emph {et~al.}(2019)\citenamefont {Arute},
  \citenamefont {Arya}, \citenamefont {Babbush}, \citenamefont {Bacon},
  \citenamefont {Bardin}, \citenamefont {Barends}, \citenamefont {Biswas},
  \citenamefont {Boixo}, \citenamefont {Brandao}, \citenamefont {Buell},
  \citenamefont {Burkett}, \citenamefont {Chen}, \citenamefont {Chen},
  \citenamefont {Chiaro}, \citenamefont {Collins}, \citenamefont {Courtney},
  \citenamefont {Dunsworth}, \citenamefont {Farhi}, \citenamefont {Foxen},
  \citenamefont {Fowler}, \citenamefont {Gidney}, \citenamefont {Giustina},
  \citenamefont {Graff}, \citenamefont {Guerin}, \citenamefont {Habegger},
  \citenamefont {Harrigan}, \citenamefont {Hartmann}, \citenamefont {Ho},
  \citenamefont {Hoffmann}, \citenamefont {Huang}, \citenamefont {Humble},
  \citenamefont {Isakov}, \citenamefont {Jeffrey}, \citenamefont {Jiang},
  \citenamefont {Kafri}, \citenamefont {Kechedzhi}, \citenamefont {Kelly},
  \citenamefont {Klimov}, \citenamefont {Knysh}, \citenamefont {Korotkov},
  \citenamefont {Kostritsa}, \citenamefont {Landhuis}, \citenamefont
  {Lindmark}, \citenamefont {Lucero}, \citenamefont {Lyakh}, \citenamefont
  {Mandr{\`a}}, \citenamefont {McClean}, \citenamefont {McEwen}, \citenamefont
  {Megrant}, \citenamefont {Mi}, \citenamefont {Michielsen}, \citenamefont
  {Mohseni}, \citenamefont {Mutus}, \citenamefont {Naaman}, \citenamefont
  {Neeley}, \citenamefont {Neill}, \citenamefont {Niu}, \citenamefont {Ostby},
  \citenamefont {Petukhov}, \citenamefont {Platt}, \citenamefont {Quintana},
  \citenamefont {Rieffel}, \citenamefont {Roushan}, \citenamefont {Rubin},
  \citenamefont {Sank}, \citenamefont {Satzinger}, \citenamefont {Smelyanskiy},
  \citenamefont {Sung}, \citenamefont {Trevithick}, \citenamefont
  {Vainsencher}, \citenamefont {Villalonga}, \citenamefont {White},
  \citenamefont {Yao}, \citenamefont {Yeh}, \citenamefont {Zalcman},
  \citenamefont {Neven},\ and\ \citenamefont {Martinis}}]{Q21}%
  \BibitemOpen
  \bibfield  {author} {\bibinfo {author} {\bibfnamefont {Frank}\ \bibnamefont
  {Arute}}, \bibinfo {author} {\bibfnamefont {Kunal}\ \bibnamefont {Arya}},
  \bibinfo {author} {\bibfnamefont {Ryan}\ \bibnamefont {Babbush}}, \bibinfo
  {author} {\bibfnamefont {Dave}\ \bibnamefont {Bacon}}, \bibinfo {author}
  {\bibfnamefont {Joseph~C.}\ \bibnamefont {Bardin}}, \bibinfo {author}
  {\bibfnamefont {Rami}\ \bibnamefont {Barends}}, \bibinfo {author}
  {\bibfnamefont {Rupak}\ \bibnamefont {Biswas}}, \bibinfo {author}
  {\bibfnamefont {Sergio}\ \bibnamefont {Boixo}}, \bibinfo {author}
  {\bibfnamefont {Fernando G. S.~L.}\ \bibnamefont {Brandao}}, \bibinfo
  {author} {\bibfnamefont {David~A.}\ \bibnamefont {Buell}}, \bibinfo {author}
  {\bibfnamefont {Brian}\ \bibnamefont {Burkett}}, \bibinfo {author}
  {\bibfnamefont {Yu}~\bibnamefont {Chen}}, \bibinfo {author} {\bibfnamefont
  {Zijun}\ \bibnamefont {Chen}}, \bibinfo {author} {\bibfnamefont {Ben}\
  \bibnamefont {Chiaro}}, \bibinfo {author} {\bibfnamefont {Roberto}\
  \bibnamefont {Collins}}, \bibinfo {author} {\bibfnamefont {William}\
  \bibnamefont {Courtney}}, \bibinfo {author} {\bibfnamefont {Andrew}\
  \bibnamefont {Dunsworth}}, \bibinfo {author} {\bibfnamefont {Edward}\
  \bibnamefont {Farhi}}, \bibinfo {author} {\bibfnamefont {Brooks}\
  \bibnamefont {Foxen}}, \bibinfo {author} {\bibfnamefont {Austin}\
  \bibnamefont {Fowler}}, \bibinfo {author} {\bibfnamefont {Craig}\
  \bibnamefont {Gidney}}, \bibinfo {author} {\bibfnamefont {Marissa}\
  \bibnamefont {Giustina}}, \bibinfo {author} {\bibfnamefont {Rob}\
  \bibnamefont {Graff}}, \bibinfo {author} {\bibfnamefont {Keith}\ \bibnamefont
  {Guerin}}, \bibinfo {author} {\bibfnamefont {Steve}\ \bibnamefont
  {Habegger}}, \bibinfo {author} {\bibfnamefont {Matthew~P.}\ \bibnamefont
  {Harrigan}}, \bibinfo {author} {\bibfnamefont {Michael~J.}\ \bibnamefont
  {Hartmann}}, \bibinfo {author} {\bibfnamefont {Alan}\ \bibnamefont {Ho}},
  \bibinfo {author} {\bibfnamefont {Markus}\ \bibnamefont {Hoffmann}}, \bibinfo
  {author} {\bibfnamefont {Trent}\ \bibnamefont {Huang}}, \bibinfo {author}
  {\bibfnamefont {Travis~S.}\ \bibnamefont {Humble}}, \bibinfo {author}
  {\bibfnamefont {Sergei~V.}\ \bibnamefont {Isakov}}, \bibinfo {author}
  {\bibfnamefont {Evan}\ \bibnamefont {Jeffrey}}, \bibinfo {author}
  {\bibfnamefont {Zhang}\ \bibnamefont {Jiang}}, \bibinfo {author}
  {\bibfnamefont {Dvir}\ \bibnamefont {Kafri}}, \bibinfo {author}
  {\bibfnamefont {Kostyantyn}\ \bibnamefont {Kechedzhi}}, \bibinfo {author}
  {\bibfnamefont {Julian}\ \bibnamefont {Kelly}}, \bibinfo {author}
  {\bibfnamefont {Paul~V.}\ \bibnamefont {Klimov}}, \bibinfo {author}
  {\bibfnamefont {Sergey}\ \bibnamefont {Knysh}}, \bibinfo {author}
  {\bibfnamefont {Alexander}\ \bibnamefont {Korotkov}}, \bibinfo {author}
  {\bibfnamefont {Fedor}\ \bibnamefont {Kostritsa}}, \bibinfo {author}
  {\bibfnamefont {David}\ \bibnamefont {Landhuis}}, \bibinfo {author}
  {\bibfnamefont {Mike}\ \bibnamefont {Lindmark}}, \bibinfo {author}
  {\bibfnamefont {Erik}\ \bibnamefont {Lucero}}, \bibinfo {author}
  {\bibfnamefont {Dmitry}\ \bibnamefont {Lyakh}}, \bibinfo {author}
  {\bibfnamefont {Salvatore}\ \bibnamefont {Mandr{\`a}}}, \bibinfo {author}
  {\bibfnamefont {Jarrod~R.}\ \bibnamefont {McClean}}, \bibinfo {author}
  {\bibfnamefont {Matthew}\ \bibnamefont {McEwen}}, \bibinfo {author}
  {\bibfnamefont {Anthony}\ \bibnamefont {Megrant}}, \bibinfo {author}
  {\bibfnamefont {Xiao}\ \bibnamefont {Mi}}, \bibinfo {author} {\bibfnamefont
  {Kristel}\ \bibnamefont {Michielsen}}, \bibinfo {author} {\bibfnamefont
  {Masoud}\ \bibnamefont {Mohseni}}, \bibinfo {author} {\bibfnamefont {Josh}\
  \bibnamefont {Mutus}}, \bibinfo {author} {\bibfnamefont {Ofer}\ \bibnamefont
  {Naaman}}, \bibinfo {author} {\bibfnamefont {Matthew}\ \bibnamefont
  {Neeley}}, \bibinfo {author} {\bibfnamefont {Charles}\ \bibnamefont {Neill}},
  \bibinfo {author} {\bibfnamefont {Murphy~Yuezhen}\ \bibnamefont {Niu}},
  \bibinfo {author} {\bibfnamefont {Eric}\ \bibnamefont {Ostby}}, \bibinfo
  {author} {\bibfnamefont {Andre}\ \bibnamefont {Petukhov}}, \bibinfo {author}
  {\bibfnamefont {John~C.}\ \bibnamefont {Platt}}, \bibinfo {author}
  {\bibfnamefont {Chris}\ \bibnamefont {Quintana}}, \bibinfo {author}
  {\bibfnamefont {Eleanor~G.}\ \bibnamefont {Rieffel}}, \bibinfo {author}
  {\bibfnamefont {Pedram}\ \bibnamefont {Roushan}}, \bibinfo {author}
  {\bibfnamefont {Nicholas~C.}\ \bibnamefont {Rubin}}, \bibinfo {author}
  {\bibfnamefont {Daniel}\ \bibnamefont {Sank}}, \bibinfo {author}
  {\bibfnamefont {Kevin~J.}\ \bibnamefont {Satzinger}}, \bibinfo {author}
  {\bibfnamefont {Vadim}\ \bibnamefont {Smelyanskiy}}, \bibinfo {author}
  {\bibfnamefont {Kevin~J.}\ \bibnamefont {Sung}}, \bibinfo {author}
  {\bibfnamefont {Matthew~D.}\ \bibnamefont {Trevithick}}, \bibinfo {author}
  {\bibfnamefont {Amit}\ \bibnamefont {Vainsencher}}, \bibinfo {author}
  {\bibfnamefont {Benjamin}\ \bibnamefont {Villalonga}}, \bibinfo {author}
  {\bibfnamefont {Theodore}\ \bibnamefont {White}}, \bibinfo {author}
  {\bibfnamefont {Z.~Jamie}\ \bibnamefont {Yao}}, \bibinfo {author}
  {\bibfnamefont {Ping}\ \bibnamefont {Yeh}}, \bibinfo {author} {\bibfnamefont
  {Adam}\ \bibnamefont {Zalcman}}, \bibinfo {author} {\bibfnamefont {Hartmut}\
  \bibnamefont {Neven}}, \ and\ \bibinfo {author} {\bibfnamefont {John~M.}\
  \bibnamefont {Martinis}},\ }\bibfield  {title} {\enquote {\bibinfo {title}
  {Quantum supremacy using a programmable superconducting processor},}\ }\href
  {\doibase 10.1038/s41586-019-1666-5} {\bibfield  {journal} {\bibinfo
  {journal} {Nature}\ }\textbf {\bibinfo {volume} {574}},\ \bibinfo {pages}
  {505--510} (\bibinfo {year} {2019})}\BibitemShut {NoStop}%
\bibitem [{\citenamefont {Preskill}(2018)}]{Q22}%
  \BibitemOpen
  \bibfield  {author} {\bibinfo {author} {\bibfnamefont {John}\ \bibnamefont
  {Preskill}},\ }\bibfield  {title} {\enquote {\bibinfo {title} {Quantum
  {C}omputing in the {NISQ} era and beyond},}\ }\href {\doibase
  10.22331/q-2018-08-06-79} {\bibfield  {journal} {\bibinfo  {journal}
  {{Quantum}}\ }\textbf {\bibinfo {volume} {2}},\ \bibinfo {pages} {79}
  (\bibinfo {year} {2018})}\BibitemShut {NoStop}%
\bibitem [{\citenamefont {Harrow}\ and\ \citenamefont {Montanaro}(2017)}]{Q23}%
  \BibitemOpen
  \bibfield  {author} {\bibinfo {author} {\bibfnamefont {Aram~W.}\ \bibnamefont
  {Harrow}}\ and\ \bibinfo {author} {\bibfnamefont {Ashley}\ \bibnamefont
  {Montanaro}},\ }\bibfield  {title} {\enquote {\bibinfo {title} {Quantum
  computational supremacy},}\ }\href {\doibase 10.1038/nature23458} {\bibfield
  {journal} {\bibinfo  {journal} {Nature}\ }\textbf {\bibinfo {volume} {549}},\
  \bibinfo {pages} {203--209} (\bibinfo {year} {2017})}\BibitemShut {NoStop}%
\bibitem [{\citenamefont {Aaronson}\ and\ \citenamefont {Chen}(2017)}]{Q24}%
  \BibitemOpen
  \bibfield  {author} {\bibinfo {author} {\bibfnamefont {Scott}\ \bibnamefont
  {Aaronson}}\ and\ \bibinfo {author} {\bibfnamefont {Lijie}\ \bibnamefont
  {Chen}},\ }\bibfield  {title} {\enquote {\bibinfo {title}
  {Complexity-theoretic foundations of quantum supremacy experiments},}\
  }\href@noop {} {\bibfield  {journal} {\bibinfo  {journal} {Proceedings of the
  32nd Computational Complexity Conference}\ ,\ \bibinfo {pages} {22:1--22:67}}
  (\bibinfo {year} {2017})}\BibitemShut {NoStop}%
\bibitem [{\citenamefont {Alexeev}\ \emph {et~al.}(2021)\citenamefont
  {Alexeev}, \citenamefont {Bacon}, \citenamefont {Brown}, \citenamefont
  {Calderbank}, \citenamefont {Carr}, \citenamefont {Chong}, \citenamefont
  {DeMarco}, \citenamefont {Englund}, \citenamefont {Farhi}, \citenamefont
  {Fefferman}, \citenamefont {Gorshkov}, \citenamefont {Houck}, \citenamefont
  {Kim}, \citenamefont {Kimmel}, \citenamefont {Lange}, \citenamefont {Lloyd},
  \citenamefont {Lukin}, \citenamefont {Maslov}, \citenamefont {Maunz},
  \citenamefont {Monroe}, \citenamefont {Preskill}, \citenamefont {Roetteler},
  \citenamefont {Savage},\ and\ \citenamefont {Thompson}}]{Q25}%
  \BibitemOpen
  \bibfield  {author} {\bibinfo {author} {\bibfnamefont {Yuri}\ \bibnamefont
  {Alexeev}}, \bibinfo {author} {\bibfnamefont {Dave}\ \bibnamefont {Bacon}},
  \bibinfo {author} {\bibfnamefont {Kenneth~R.}\ \bibnamefont {Brown}},
  \bibinfo {author} {\bibfnamefont {Robert}\ \bibnamefont {Calderbank}},
  \bibinfo {author} {\bibfnamefont {Lincoln~D.}\ \bibnamefont {Carr}}, \bibinfo
  {author} {\bibfnamefont {Frederic~T.}\ \bibnamefont {Chong}}, \bibinfo
  {author} {\bibfnamefont {Brian}\ \bibnamefont {DeMarco}}, \bibinfo {author}
  {\bibfnamefont {Dirk}\ \bibnamefont {Englund}}, \bibinfo {author}
  {\bibfnamefont {Edward}\ \bibnamefont {Farhi}}, \bibinfo {author}
  {\bibfnamefont {Bill}\ \bibnamefont {Fefferman}}, \bibinfo {author}
  {\bibfnamefont {Alexey~V.}\ \bibnamefont {Gorshkov}}, \bibinfo {author}
  {\bibfnamefont {Andrew}\ \bibnamefont {Houck}}, \bibinfo {author}
  {\bibfnamefont {Jungsang}\ \bibnamefont {Kim}}, \bibinfo {author}
  {\bibfnamefont {Shelby}\ \bibnamefont {Kimmel}}, \bibinfo {author}
  {\bibfnamefont {Michael}\ \bibnamefont {Lange}}, \bibinfo {author}
  {\bibfnamefont {Seth}\ \bibnamefont {Lloyd}}, \bibinfo {author}
  {\bibfnamefont {Mikhail~D.}\ \bibnamefont {Lukin}}, \bibinfo {author}
  {\bibfnamefont {Dmitri}\ \bibnamefont {Maslov}}, \bibinfo {author}
  {\bibfnamefont {Peter}\ \bibnamefont {Maunz}}, \bibinfo {author}
  {\bibfnamefont {Christopher}\ \bibnamefont {Monroe}}, \bibinfo {author}
  {\bibfnamefont {John}\ \bibnamefont {Preskill}}, \bibinfo {author}
  {\bibfnamefont {Martin}\ \bibnamefont {Roetteler}}, \bibinfo {author}
  {\bibfnamefont {Martin~J.}\ \bibnamefont {Savage}}, \ and\ \bibinfo {author}
  {\bibfnamefont {Jeff}\ \bibnamefont {Thompson}},\ }\bibfield  {title}
  {\enquote {\bibinfo {title} {Quantum computer systems for scientific
  discovery},}\ }\href {\doibase 10.1103/PRXQuantum.2.017001} {\bibfield
  {journal} {\bibinfo  {journal} {PRX Quantum}\ }\textbf {\bibinfo {volume}
  {2}},\ \bibinfo {pages} {017001} (\bibinfo {year} {2021})}\BibitemShut
  {NoStop}%
\bibitem [{\citenamefont {Awschalom}\ \emph {et~al.}(2021)\citenamefont
  {Awschalom}, \citenamefont {Berggren}, \citenamefont {Bernien}, \citenamefont
  {Bhave}, \citenamefont {Carr}, \citenamefont {Davids}, \citenamefont
  {Economou}, \citenamefont {Englund}, \citenamefont {Faraon}, \citenamefont
  {Fejer}, \citenamefont {Guha}, \citenamefont {Gustafsson}, \citenamefont
  {Hu}, \citenamefont {Jiang}, \citenamefont {Kim}, \citenamefont {Korzh},
  \citenamefont {Kumar}, \citenamefont {Kwiat}, \citenamefont
  {Lon\ifmmode~\check{c}\else \v{c}\fi{}ar}, \citenamefont {Lukin},
  \citenamefont {Miller}, \citenamefont {Monroe}, \citenamefont {Nam},
  \citenamefont {Narang}, \citenamefont {Orcutt}, \citenamefont {Raymer},
  \citenamefont {Safavi-Naeini}, \citenamefont {Spiropulu}, \citenamefont
  {Srinivasan}, \citenamefont {Sun}, \citenamefont {Vu\ifmmode \check{c}\else
  \v{c}\fi{}kovi\ifmmode~\acute{c}\else \'{c}\fi{}}, \citenamefont {Waks},
  \citenamefont {Walsworth}, \citenamefont {Weiner},\ and\ \citenamefont
  {Zhang}}]{Q26}%
  \BibitemOpen
  \bibfield  {author} {\bibinfo {author} {\bibfnamefont {David}\ \bibnamefont
  {Awschalom}}, \bibinfo {author} {\bibfnamefont {Karl~K.}\ \bibnamefont
  {Berggren}}, \bibinfo {author} {\bibfnamefont {Hannes}\ \bibnamefont
  {Bernien}}, \bibinfo {author} {\bibfnamefont {Sunil}\ \bibnamefont {Bhave}},
  \bibinfo {author} {\bibfnamefont {Lincoln~D.}\ \bibnamefont {Carr}}, \bibinfo
  {author} {\bibfnamefont {Paul}\ \bibnamefont {Davids}}, \bibinfo {author}
  {\bibfnamefont {Sophia~E.}\ \bibnamefont {Economou}}, \bibinfo {author}
  {\bibfnamefont {Dirk}\ \bibnamefont {Englund}}, \bibinfo {author}
  {\bibfnamefont {Andrei}\ \bibnamefont {Faraon}}, \bibinfo {author}
  {\bibfnamefont {Martin}\ \bibnamefont {Fejer}}, \bibinfo {author}
  {\bibfnamefont {Saikat}\ \bibnamefont {Guha}}, \bibinfo {author}
  {\bibfnamefont {Martin~V.}\ \bibnamefont {Gustafsson}}, \bibinfo {author}
  {\bibfnamefont {Evelyn}\ \bibnamefont {Hu}}, \bibinfo {author} {\bibfnamefont
  {Liang}\ \bibnamefont {Jiang}}, \bibinfo {author} {\bibfnamefont {Jungsang}\
  \bibnamefont {Kim}}, \bibinfo {author} {\bibfnamefont {Boris}\ \bibnamefont
  {Korzh}}, \bibinfo {author} {\bibfnamefont {Prem}\ \bibnamefont {Kumar}},
  \bibinfo {author} {\bibfnamefont {Paul~G.}\ \bibnamefont {Kwiat}}, \bibinfo
  {author} {\bibfnamefont {Marko}\ \bibnamefont {Lon\ifmmode~\check{c}\else
  \v{c}\fi{}ar}}, \bibinfo {author} {\bibfnamefont {Mikhail~D.}\ \bibnamefont
  {Lukin}}, \bibinfo {author} {\bibfnamefont {David~A.B.}\ \bibnamefont
  {Miller}}, \bibinfo {author} {\bibfnamefont {Christopher}\ \bibnamefont
  {Monroe}}, \bibinfo {author} {\bibfnamefont {Sae~Woo}\ \bibnamefont {Nam}},
  \bibinfo {author} {\bibfnamefont {Prineha}\ \bibnamefont {Narang}}, \bibinfo
  {author} {\bibfnamefont {Jason~S.}\ \bibnamefont {Orcutt}}, \bibinfo {author}
  {\bibfnamefont {Michael~G.}\ \bibnamefont {Raymer}}, \bibinfo {author}
  {\bibfnamefont {Amir~H.}\ \bibnamefont {Safavi-Naeini}}, \bibinfo {author}
  {\bibfnamefont {Maria}\ \bibnamefont {Spiropulu}}, \bibinfo {author}
  {\bibfnamefont {Kartik}\ \bibnamefont {Srinivasan}}, \bibinfo {author}
  {\bibfnamefont {Shuo}\ \bibnamefont {Sun}}, \bibinfo {author} {\bibfnamefont
  {Jelena}\ \bibnamefont {Vu\ifmmode \check{c}\else
  \v{c}\fi{}kovi\ifmmode~\acute{c}\else \'{c}\fi{}}}, \bibinfo {author}
  {\bibfnamefont {Edo}\ \bibnamefont {Waks}}, \bibinfo {author} {\bibfnamefont
  {Ronald}\ \bibnamefont {Walsworth}}, \bibinfo {author} {\bibfnamefont
  {Andrew~M.}\ \bibnamefont {Weiner}}, \ and\ \bibinfo {author} {\bibfnamefont
  {Zheshen}\ \bibnamefont {Zhang}},\ }\bibfield  {title} {\enquote {\bibinfo
  {title} {Development of quantum interconnects (quics) for next-generation
  information technologies},}\ }\href {\doibase 10.1103/PRXQuantum.2.017002}
  {\bibfield  {journal} {\bibinfo  {journal} {PRX Quantum}\ }\textbf {\bibinfo
  {volume} {2}},\ \bibinfo {pages} {017002} (\bibinfo {year}
  {2021})}\BibitemShut {NoStop}%
\bibitem [{\citenamefont {Foxen}\ \emph {et~al.}(2020)\citenamefont {Foxen},
  \citenamefont {Neill}, \citenamefont {Dunsworth}, \citenamefont {Roushan},
  \citenamefont {Chiaro}, \citenamefont {Megrant}, \citenamefont {Kelly},
  \citenamefont {Chen}, \citenamefont {Satzinger}, \citenamefont {Barends},
  \citenamefont {Arute}, \citenamefont {Arya}, \citenamefont {Babbush},
  \citenamefont {Bacon}, \citenamefont {Bardin}, \citenamefont {Boixo},
  \citenamefont {Buell}, \citenamefont {Burkett}, \citenamefont {Chen},
  \citenamefont {Collins}, \citenamefont {Farhi}, \citenamefont {Fowler},
  \citenamefont {Gidney}, \citenamefont {Giustina}, \citenamefont {Graff},
  \citenamefont {Harrigan}, \citenamefont {Huang}, \citenamefont {Isakov},
  \citenamefont {Jeffrey}, \citenamefont {Jiang}, \citenamefont {Kafri},
  \citenamefont {Kechedzhi}, \citenamefont {Klimov}, \citenamefont {Korotkov},
  \citenamefont {Kostritsa}, \citenamefont {Landhuis}, \citenamefont {Lucero},
  \citenamefont {McClean}, \citenamefont {McEwen}, \citenamefont {Mi},
  \citenamefont {Mohseni}, \citenamefont {Mutus}, \citenamefont {Naaman},
  \citenamefont {Neeley}, \citenamefont {Niu}, \citenamefont {Petukhov},
  \citenamefont {Quintana}, \citenamefont {Rubin}, \citenamefont {Sank},
  \citenamefont {Smelyanskiy}, \citenamefont {Vainsencher}, \citenamefont
  {White}, \citenamefont {Yao}, \citenamefont {Yeh}, \citenamefont {Zalcman},
  \citenamefont {Neven},\ and\ \citenamefont {Martinis}}]{Q27}%
  \BibitemOpen
  \bibfield  {author} {\bibinfo {author} {\bibfnamefont {B.}~\bibnamefont
  {Foxen}}, \bibinfo {author} {\bibfnamefont {C.}~\bibnamefont {Neill}},
  \bibinfo {author} {\bibfnamefont {A.}~\bibnamefont {Dunsworth}}, \bibinfo
  {author} {\bibfnamefont {P.}~\bibnamefont {Roushan}}, \bibinfo {author}
  {\bibfnamefont {B.}~\bibnamefont {Chiaro}}, \bibinfo {author} {\bibfnamefont
  {A.}~\bibnamefont {Megrant}}, \bibinfo {author} {\bibfnamefont
  {J.}~\bibnamefont {Kelly}}, \bibinfo {author} {\bibfnamefont {Zijun}\
  \bibnamefont {Chen}}, \bibinfo {author} {\bibfnamefont {K.}~\bibnamefont
  {Satzinger}}, \bibinfo {author} {\bibfnamefont {R.}~\bibnamefont {Barends}},
  \bibinfo {author} {\bibfnamefont {F.}~\bibnamefont {Arute}}, \bibinfo
  {author} {\bibfnamefont {K.}~\bibnamefont {Arya}}, \bibinfo {author}
  {\bibfnamefont {R.}~\bibnamefont {Babbush}}, \bibinfo {author} {\bibfnamefont
  {D.}~\bibnamefont {Bacon}}, \bibinfo {author} {\bibfnamefont {J.~C.}\
  \bibnamefont {Bardin}}, \bibinfo {author} {\bibfnamefont {S.}~\bibnamefont
  {Boixo}}, \bibinfo {author} {\bibfnamefont {D.}~\bibnamefont {Buell}},
  \bibinfo {author} {\bibfnamefont {B.}~\bibnamefont {Burkett}}, \bibinfo
  {author} {\bibfnamefont {Yu}~\bibnamefont {Chen}}, \bibinfo {author}
  {\bibfnamefont {R.}~\bibnamefont {Collins}}, \bibinfo {author} {\bibfnamefont
  {E.}~\bibnamefont {Farhi}}, \bibinfo {author} {\bibfnamefont
  {A.}~\bibnamefont {Fowler}}, \bibinfo {author} {\bibfnamefont
  {C.}~\bibnamefont {Gidney}}, \bibinfo {author} {\bibfnamefont
  {M.}~\bibnamefont {Giustina}}, \bibinfo {author} {\bibfnamefont
  {R.}~\bibnamefont {Graff}}, \bibinfo {author} {\bibfnamefont
  {M.}~\bibnamefont {Harrigan}}, \bibinfo {author} {\bibfnamefont
  {T.}~\bibnamefont {Huang}}, \bibinfo {author} {\bibfnamefont {S.~V.}\
  \bibnamefont {Isakov}}, \bibinfo {author} {\bibfnamefont {E.}~\bibnamefont
  {Jeffrey}}, \bibinfo {author} {\bibfnamefont {Z.}~\bibnamefont {Jiang}},
  \bibinfo {author} {\bibfnamefont {D.}~\bibnamefont {Kafri}}, \bibinfo
  {author} {\bibfnamefont {K.}~\bibnamefont {Kechedzhi}}, \bibinfo {author}
  {\bibfnamefont {P.}~\bibnamefont {Klimov}}, \bibinfo {author} {\bibfnamefont
  {A.}~\bibnamefont {Korotkov}}, \bibinfo {author} {\bibfnamefont
  {F.}~\bibnamefont {Kostritsa}}, \bibinfo {author} {\bibfnamefont
  {D.}~\bibnamefont {Landhuis}}, \bibinfo {author} {\bibfnamefont
  {E.}~\bibnamefont {Lucero}}, \bibinfo {author} {\bibfnamefont
  {J.}~\bibnamefont {McClean}}, \bibinfo {author} {\bibfnamefont
  {M.}~\bibnamefont {McEwen}}, \bibinfo {author} {\bibfnamefont
  {X.}~\bibnamefont {Mi}}, \bibinfo {author} {\bibfnamefont {M.}~\bibnamefont
  {Mohseni}}, \bibinfo {author} {\bibfnamefont {J.~Y.}\ \bibnamefont {Mutus}},
  \bibinfo {author} {\bibfnamefont {O.}~\bibnamefont {Naaman}}, \bibinfo
  {author} {\bibfnamefont {M.}~\bibnamefont {Neeley}}, \bibinfo {author}
  {\bibfnamefont {M.}~\bibnamefont {Niu}}, \bibinfo {author} {\bibfnamefont
  {A.}~\bibnamefont {Petukhov}}, \bibinfo {author} {\bibfnamefont
  {C.}~\bibnamefont {Quintana}}, \bibinfo {author} {\bibfnamefont
  {N.}~\bibnamefont {Rubin}}, \bibinfo {author} {\bibfnamefont
  {D.}~\bibnamefont {Sank}}, \bibinfo {author} {\bibfnamefont {V.}~\bibnamefont
  {Smelyanskiy}}, \bibinfo {author} {\bibfnamefont {A.}~\bibnamefont
  {Vainsencher}}, \bibinfo {author} {\bibfnamefont {T.~C.}\ \bibnamefont
  {White}}, \bibinfo {author} {\bibfnamefont {Z.}~\bibnamefont {Yao}}, \bibinfo
  {author} {\bibfnamefont {P.}~\bibnamefont {Yeh}}, \bibinfo {author}
  {\bibfnamefont {A.}~\bibnamefont {Zalcman}}, \bibinfo {author} {\bibfnamefont
  {H.}~\bibnamefont {Neven}}, \ and\ \bibinfo {author} {\bibfnamefont {J.~M.}\
  \bibnamefont {Martinis}} (\bibinfo {collaboration} {Google AI Quantum}),\
  }\bibfield  {title} {\enquote {\bibinfo {title} {Demonstrating a continuous
  set of two-qubit gates for near-term quantum algorithms},}\ }\href {\doibase
  10.1103/PhysRevLett.125.120504} {\bibfield  {journal} {\bibinfo  {journal}
  {Phys. Rev. Lett.}\ }\textbf {\bibinfo {volume} {125}},\ \bibinfo {pages}
  {120504} (\bibinfo {year} {2020})}\BibitemShut {NoStop}%
\bibitem [{\citenamefont {Pirandola}\ and\ \citenamefont
  {Braunstein}(2016)}]{Q31}%
  \BibitemOpen
  \bibfield  {author} {\bibinfo {author} {\bibfnamefont {Stefano}\ \bibnamefont
  {Pirandola}}\ and\ \bibinfo {author} {\bibfnamefont {Samuel~L.}\ \bibnamefont
  {Braunstein}},\ }\bibfield  {title} {\enquote {\bibinfo {title} {Physics:
  Unite to build a quantum internet},}\ }\href {\doibase 10.1038/532169a}
  {\bibfield  {journal} {\bibinfo  {journal} {Nature}\ }\textbf {\bibinfo
  {volume} {532}},\ \bibinfo {pages} {169--171} (\bibinfo {year}
  {2016})}\BibitemShut {NoStop}%
\bibitem [{\citenamefont {Wehner}\ \emph {et~al.}(2018)\citenamefont {Wehner},
  \citenamefont {Elkouss},\ and\ \citenamefont {Hanson}}]{Q32}%
  \BibitemOpen
  \bibfield  {author} {\bibinfo {author} {\bibfnamefont {Stephanie}\
  \bibnamefont {Wehner}}, \bibinfo {author} {\bibfnamefont {David}\
  \bibnamefont {Elkouss}}, \ and\ \bibinfo {author} {\bibfnamefont {Ronald}\
  \bibnamefont {Hanson}},\ }\bibfield  {title} {\enquote {\bibinfo {title}
  {Quantum internet: A vision for the road ahead},}\ }\href {\doibase
  10.1126/science.aam9288} {\bibfield  {journal} {\bibinfo  {journal}
  {Science}\ }\textbf {\bibinfo {volume} {362}} (\bibinfo {year} {2018}),\
  10.1126/science.aam9288},\ \Eprint
  {http://arxiv.org/abs/https://science.sciencemag.org/content/362/6412/eaam9288.full.pdf}
  {https://science.sciencemag.org/content/362/6412/eaam9288.full.pdf}
  \BibitemShut {NoStop}%
\bibitem [{\citenamefont {Pirandola}\ \emph {et~al.}(2020)\citenamefont
  {Pirandola}, \citenamefont {Andersen}, \citenamefont {Banchi}, \citenamefont
  {Berta}, \citenamefont {Bunandar}, \citenamefont {Colbeck}, \citenamefont
  {Englund}, \citenamefont {Gehring}, \citenamefont {Lupo}, \citenamefont
  {Ottaviani}, \citenamefont {Pereira}, \citenamefont {Razavi}, \citenamefont
  {Shaari}, \citenamefont {Tomamichel}, \citenamefont {Usenko}, \citenamefont
  {Vallone}, \citenamefont {Villoresi},\ and\ \citenamefont {Wallden}}]{Q33}%
  \BibitemOpen
  \bibfield  {author} {\bibinfo {author} {\bibfnamefont {S.}~\bibnamefont
  {Pirandola}}, \bibinfo {author} {\bibfnamefont {U.~L.}\ \bibnamefont
  {Andersen}}, \bibinfo {author} {\bibfnamefont {L.}~\bibnamefont {Banchi}},
  \bibinfo {author} {\bibfnamefont {M.}~\bibnamefont {Berta}}, \bibinfo
  {author} {\bibfnamefont {D.}~\bibnamefont {Bunandar}}, \bibinfo {author}
  {\bibfnamefont {R.}~\bibnamefont {Colbeck}}, \bibinfo {author} {\bibfnamefont
  {D.}~\bibnamefont {Englund}}, \bibinfo {author} {\bibfnamefont
  {T.}~\bibnamefont {Gehring}}, \bibinfo {author} {\bibfnamefont
  {C.}~\bibnamefont {Lupo}}, \bibinfo {author} {\bibfnamefont {C.}~\bibnamefont
  {Ottaviani}}, \bibinfo {author} {\bibfnamefont {J.~L.}\ \bibnamefont
  {Pereira}}, \bibinfo {author} {\bibfnamefont {M.}~\bibnamefont {Razavi}},
  \bibinfo {author} {\bibfnamefont {J.~Shamsul}\ \bibnamefont {Shaari}},
  \bibinfo {author} {\bibfnamefont {M.}~\bibnamefont {Tomamichel}}, \bibinfo
  {author} {\bibfnamefont {V.~C.}\ \bibnamefont {Usenko}}, \bibinfo {author}
  {\bibfnamefont {G.}~\bibnamefont {Vallone}}, \bibinfo {author} {\bibfnamefont
  {P.}~\bibnamefont {Villoresi}}, \ and\ \bibinfo {author} {\bibfnamefont
  {P.}~\bibnamefont {Wallden}},\ }\bibfield  {title} {\enquote {\bibinfo
  {title} {Advances in quantum cryptography},}\ }\href {\doibase
  10.1364/AOP.361502} {\bibfield  {journal} {\bibinfo  {journal} {Adv. Opt.
  Photon.}\ }\textbf {\bibinfo {volume} {12}},\ \bibinfo {pages} {1012--1236}
  (\bibinfo {year} {2020})}\BibitemShut {NoStop}%
\bibitem [{\citenamefont {Pirandola}(2019{\natexlab{a}})}]{Q34}%
  \BibitemOpen
  \bibfield  {author} {\bibinfo {author} {\bibfnamefont {Stefano}\ \bibnamefont
  {Pirandola}},\ }\bibfield  {title} {\enquote {\bibinfo {title} {End-to-end
  capacities of a quantum communication network},}\ }\href {\doibase
  10.1038/s42005-019-0147-3} {\bibfield  {journal} {\bibinfo  {journal}
  {Communications Physics}\ }\textbf {\bibinfo {volume} {2}},\ \bibinfo {pages}
  {51} (\bibinfo {year} {2019}{\natexlab{a}})}\BibitemShut {NoStop}%
\bibitem [{\citenamefont {Pirandola}\ \emph {et~al.}(2017)\citenamefont
  {Pirandola}, \citenamefont {Laurenza}, \citenamefont {Ottaviani},\ and\
  \citenamefont {Banchi}}]{Q35}%
  \BibitemOpen
  \bibfield  {author} {\bibinfo {author} {\bibfnamefont {Stefano}\ \bibnamefont
  {Pirandola}}, \bibinfo {author} {\bibfnamefont {Riccardo}\ \bibnamefont
  {Laurenza}}, \bibinfo {author} {\bibfnamefont {Carlo}\ \bibnamefont
  {Ottaviani}}, \ and\ \bibinfo {author} {\bibfnamefont {Leonardo}\
  \bibnamefont {Banchi}},\ }\bibfield  {title} {\enquote {\bibinfo {title}
  {Fundamental limits of repeaterless quantum communications},}\ }\href
  {\doibase 10.1038/ncomms15043} {\bibfield  {journal} {\bibinfo  {journal}
  {Nature Communications}\ }\textbf {\bibinfo {volume} {8}},\ \bibinfo {pages}
  {15043} (\bibinfo {year} {2017})}\BibitemShut {NoStop}%
\bibitem [{\citenamefont {Pirandola}\ \emph {et~al.}(2018)\citenamefont
  {Pirandola}, \citenamefont {Braunstein}, \citenamefont {Laurenza},
  \citenamefont {Ottaviani}, \citenamefont {Cope}, \citenamefont {Spedalieri},\
  and\ \citenamefont {Banchi}}]{Q36}%
  \BibitemOpen
  \bibfield  {author} {\bibinfo {author} {\bibfnamefont {Stefano}\ \bibnamefont
  {Pirandola}}, \bibinfo {author} {\bibfnamefont {Samuel~L}\ \bibnamefont
  {Braunstein}}, \bibinfo {author} {\bibfnamefont {Riccardo}\ \bibnamefont
  {Laurenza}}, \bibinfo {author} {\bibfnamefont {Carlo}\ \bibnamefont
  {Ottaviani}}, \bibinfo {author} {\bibfnamefont {Thomas P~W}\ \bibnamefont
  {Cope}}, \bibinfo {author} {\bibfnamefont {Gaetana}\ \bibnamefont
  {Spedalieri}}, \ and\ \bibinfo {author} {\bibfnamefont {Leonardo}\
  \bibnamefont {Banchi}},\ }\bibfield  {title} {\enquote {\bibinfo {title}
  {Theory of channel simulation and bounds for private communication},}\ }\href
  {\doibase 10.1088/2058-9565/aac394} {\bibfield  {journal} {\bibinfo
  {journal} {Quantum Science and Technology}\ }\textbf {\bibinfo {volume}
  {3}},\ \bibinfo {pages} {035009} (\bibinfo {year} {2018})}\BibitemShut
  {NoStop}%
\bibitem [{\citenamefont {Pirandola}(2019{\natexlab{b}})}]{Q37}%
  \BibitemOpen
  \bibfield  {author} {\bibinfo {author} {\bibfnamefont {Stefano}\ \bibnamefont
  {Pirandola}},\ }\bibfield  {title} {\enquote {\bibinfo {title} {Bounds for
  multi-end communication over quantum networks},}\ }\href {\doibase
  10.1088/2058-9565/ab3f66} {\bibfield  {journal} {\bibinfo  {journal} {Quantum
  Science and Technology}\ }\textbf {\bibinfo {volume} {4}},\ \bibinfo {pages}
  {045006} (\bibinfo {year} {2019}{\natexlab{b}})}\BibitemShut {NoStop}%
\bibitem [{\citenamefont {MacQuarrie}\ \emph
  {et~al.}(2020{\natexlab{b}})\citenamefont {MacQuarrie}, \citenamefont
  {Simon}, \citenamefont {Simmons},\ and\ \citenamefont
  {Maine}}]{MacQuarrie2020}%
  \BibitemOpen
  \bibfield  {author} {\bibinfo {author} {\bibfnamefont {Evan~R.}\ \bibnamefont
  {MacQuarrie}}, \bibinfo {author} {\bibfnamefont {Christoph}\ \bibnamefont
  {Simon}}, \bibinfo {author} {\bibfnamefont {Stephanie}\ \bibnamefont
  {Simmons}}, \ and\ \bibinfo {author} {\bibfnamefont {Elicia}\ \bibnamefont
  {Maine}},\ }\bibfield  {title} {\enquote {\bibinfo {title} {The emerging
  commercial landscape of quantum computing},}\ }\href {\doibase
  10.1038/s42254-020-00247-5} {\bibfield  {journal} {\bibinfo  {journal}
  {Nature Reviews Physics}\ }\textbf {\bibinfo {volume} {2}},\ \bibinfo {pages}
  {596--598} (\bibinfo {year} {2020}{\natexlab{b}})}\BibitemShut {NoStop}%
\bibitem [{\citenamefont {Or{\'{u}}s}\ \emph {et~al.}(2019)\citenamefont
  {Or{\'{u}}s}, \citenamefont {Mugel},\ and\ \citenamefont
  {Lizaso}}]{Orus2018}%
  \BibitemOpen
  \bibfield  {author} {\bibinfo {author} {\bibfnamefont {Rom{\'{a}}n}\
  \bibnamefont {Or{\'{u}}s}}, \bibinfo {author} {\bibfnamefont {Samuel}\
  \bibnamefont {Mugel}}, \ and\ \bibinfo {author} {\bibfnamefont {Enrique}\
  \bibnamefont {Lizaso}},\ }\bibfield  {title} {\enquote {\bibinfo {title}
  {{Quantum computing for finance: Overview and prospects}},}\ }\href {\doibase
  10.1016/j.revip.2019.100028} {\bibfield  {journal} {\bibinfo  {journal}
  {Reviews in Physics}\ }\textbf {\bibinfo {volume} {4}},\ \bibinfo {pages}
  {100028} (\bibinfo {year} {2019})},\ \Eprint
  {http://arxiv.org/abs/1807.03890} {arXiv:1807.03890} \BibitemShut {NoStop}%
\bibitem [{\citenamefont {Mugel}\ \emph
  {et~al.}(2020{\natexlab{a}})\citenamefont {Mugel}, \citenamefont {Lizaso},\
  and\ \citenamefont {Orus}}]{Mugel2020a}%
  \BibitemOpen
  \bibfield  {author} {\bibinfo {author} {\bibfnamefont {Samuel}\ \bibnamefont
  {Mugel}}, \bibinfo {author} {\bibfnamefont {Enrique}\ \bibnamefont {Lizaso}},
  \ and\ \bibinfo {author} {\bibfnamefont {Roman}\ \bibnamefont {Orus}},\
  }\bibfield  {title} {\enquote {\bibinfo {title} {{Use Cases of Quantum
  Optimization for Finance}},}\ }\href {http://arxiv.org/abs/2010.01312}
  {\bibfield  {journal} {\bibinfo  {journal} {arXiv:2010.01312}\ ,\ \bibinfo
  {pages} {1--10}} (\bibinfo {year} {2020}{\natexlab{a}})},\ \Eprint
  {http://arxiv.org/abs/2010.01312} {arXiv:2010.01312} \BibitemShut {NoStop}%
\bibitem [{\citenamefont {Rosenberg}\ \emph {et~al.}(2016)\citenamefont
  {Rosenberg}, \citenamefont {Haghnegahdar}, \citenamefont {Goddard},
  \citenamefont {Carr}, \citenamefont {Wu},\ and\ \citenamefont {{L{\'{o}}pez
  De Prado}}}]{Rosenberg2016}%
  \BibitemOpen
  \bibfield  {author} {\bibinfo {author} {\bibfnamefont {Gili}\ \bibnamefont
  {Rosenberg}}, \bibinfo {author} {\bibfnamefont {Poya}\ \bibnamefont
  {Haghnegahdar}}, \bibinfo {author} {\bibfnamefont {Phil}\ \bibnamefont
  {Goddard}}, \bibinfo {author} {\bibfnamefont {Peter}\ \bibnamefont {Carr}},
  \bibinfo {author} {\bibfnamefont {Kesheng}\ \bibnamefont {Wu}}, \ and\
  \bibinfo {author} {\bibfnamefont {Marcos}\ \bibnamefont {{L{\'{o}}pez De
  Prado}}},\ }\bibfield  {title} {\enquote {\bibinfo {title} {{Solving the
  Optimal Trading Trajectory Problem Using a Quantum Annealer}},}\ }\href
  {\doibase 10.1109/JSTSP.2016.2574703} {\bibfield  {journal} {\bibinfo
  {journal} {IEEE Journal of Selected Topics in Signal Processing}\ }\textbf
  {\bibinfo {volume} {10}},\ \bibinfo {pages} {1053--1060} (\bibinfo {year}
  {2016})}\BibitemShut {NoStop}%
\bibitem [{\citenamefont {Elsokkary}\ \emph {et~al.}(2017)\citenamefont
  {Elsokkary}, \citenamefont {Khan}, \citenamefont {Torre}, \citenamefont
  {Humble},\ and\ \citenamefont {Gottlieb}}]{Elsokkary2017a}%
  \BibitemOpen
  \bibfield  {author} {\bibinfo {author} {\bibfnamefont {Nada}\ \bibnamefont
  {Elsokkary}}, \bibinfo {author} {\bibfnamefont {Faisal~Shah}\ \bibnamefont
  {Khan}}, \bibinfo {author} {\bibfnamefont {Davide~La}\ \bibnamefont {Torre}},
  \bibinfo {author} {\bibfnamefont {Travis~S}\ \bibnamefont {Humble}}, \ and\
  \bibinfo {author} {\bibfnamefont {Joel}\ \bibnamefont {Gottlieb}},\
  }\bibfield  {title} {\enquote {\bibinfo {title} {{Financial Portfolio
  Management using D-Wave's Quantum Optimizer: The Case of Abu Dhabi Securities
  Exchange}},}\ }\href
  {http://ieee-hpec.org/2017/techprog2017/index_htm_files/102.pdf} {\bibfield
  {journal} {\bibinfo  {journal} {Ieee}\ ,\ \bibinfo {pages} {1--4}} (\bibinfo
  {year} {2017})}\BibitemShut {NoStop}%
\bibitem [{\citenamefont {Grant}\ \emph {et~al.}(2020)\citenamefont {Grant},
  \citenamefont {Humble},\ and\ \citenamefont {Stump}}]{Grant2020}%
  \BibitemOpen
  \bibfield  {author} {\bibinfo {author} {\bibfnamefont {Erica}\ \bibnamefont
  {Grant}}, \bibinfo {author} {\bibfnamefont {Travis}\ \bibnamefont {Humble}},
  \ and\ \bibinfo {author} {\bibfnamefont {Benjamin}\ \bibnamefont {Stump}},\
  }\bibfield  {title} {\enquote {\bibinfo {title} {{Benchmarking Quantum
  Annealing Controls with Portfolio Optimization}},}\ }\href
  {http://arxiv.org/abs/2007.03005} {\bibfield  {journal} {\bibinfo  {journal}
  {arXiv:2007.03005}\ } (\bibinfo {year} {2020})},\ \Eprint
  {http://arxiv.org/abs/2007.03005} {arXiv:2007.03005} \BibitemShut {NoStop}%
\bibitem [{\citenamefont {Cohen}\ \emph {et~al.}(2020)\citenamefont {Cohen},
  \citenamefont {Khan},\ and\ \citenamefont {Alexander}}]{Cohen2020}%
  \BibitemOpen
  \bibfield  {author} {\bibinfo {author} {\bibfnamefont {Jeffrey}\ \bibnamefont
  {Cohen}}, \bibinfo {author} {\bibfnamefont {Alex}\ \bibnamefont {Khan}}, \
  and\ \bibinfo {author} {\bibfnamefont {Clark}\ \bibnamefont {Alexander}},\
  }\bibfield  {title} {\enquote {\bibinfo {title} {{Portfolio Optimization of
  40 Stocks Using the DWave Quantum Annealer}},}\ }\href
  {http://arxiv.org/abs/2007.01430} {\bibfield  {journal} {\bibinfo  {journal}
  {arXiv:2007.01430}\ ,\ \bibinfo {pages} {1--16}} (\bibinfo {year} {2020})},\
  \Eprint {http://arxiv.org/abs/2007.01430} {arXiv:2007.01430} \BibitemShut
  {NoStop}%
\bibitem [{\citenamefont {Mugel}\ \emph
  {et~al.}(2020{\natexlab{b}})\citenamefont {Mugel}, \citenamefont
  {Kuchkovsky}, \citenamefont {Sanchez}, \citenamefont {Fernandez-Lorenzo},
  \citenamefont {Luis-Hita}, \citenamefont {Lizaso},\ and\ \citenamefont
  {Orus}}]{Mugel2020}%
  \BibitemOpen
  \bibfield  {author} {\bibinfo {author} {\bibfnamefont {Samuel}\ \bibnamefont
  {Mugel}}, \bibinfo {author} {\bibfnamefont {Carlos}\ \bibnamefont
  {Kuchkovsky}}, \bibinfo {author} {\bibfnamefont {Escolastico}\ \bibnamefont
  {Sanchez}}, \bibinfo {author} {\bibfnamefont {Samuel}\ \bibnamefont
  {Fernandez-Lorenzo}}, \bibinfo {author} {\bibfnamefont {Jorge}\ \bibnamefont
  {Luis-Hita}}, \bibinfo {author} {\bibfnamefont {Enrique}\ \bibnamefont
  {Lizaso}}, \ and\ \bibinfo {author} {\bibfnamefont {Roman}\ \bibnamefont
  {Orus}},\ }\bibfield  {title} {\enquote {\bibinfo {title} {{Dynamic Portfolio
  Optimization with Real Datasets Using Quantum Processors and Quantum-Inspired
  Tensor Networks}},}\ }\href {http://arxiv.org/abs/2007.00017} {\bibfield
  {journal} {\bibinfo  {journal} {arXiv:2007.00017}\ ,\ \bibinfo {pages}
  {1--11}} (\bibinfo {year} {2020}{\natexlab{b}})},\ \Eprint
  {http://arxiv.org/abs/2007.00017} {arXiv:2007.00017} \BibitemShut {NoStop}%
\bibitem [{\citenamefont {Markowitz}(1952)}]{Singleton2018}%
  \BibitemOpen
  \bibfield  {author} {\bibinfo {author} {\bibfnamefont {Harry}\ \bibnamefont
  {Markowitz}},\ }\bibfield  {title} {\enquote {\bibinfo {title} {{Portfolio
  Selection}},}\ }\href {\doibase 10.2307/2975974} {\bibfield  {journal}
  {\bibinfo  {journal} {The Journal of Finance}\ }\textbf {\bibinfo {volume}
  {7}},\ \bibinfo {pages} {77} (\bibinfo {year} {1952})}\BibitemShut {NoStop}%
\end{thebibliography}%

\end{document}